%% file: main.tex
\documentclass[letterpaper]{article} 
\usepackage{aaai2026}  
\usepackage{times}  
\usepackage{helvet}  
\usepackage{courier}  
\usepackage[hyphens]{url}  
\usepackage{graphicx} 
\usepackage{natbib}  
\usepackage{caption} 
\usepackage{algorithm}
\usepackage{algorithmic}

\usepackage{newfloat}
\usepackage{listings}
\DeclareCaptionStyle{ruled}{labelfont=normalfont,labelsep=colon,strut=off} 
\floatstyle{ruled}
\newfloat{listing}{tb}{lst}{}
\floatname{listing}{Listing}
\usepackage{amssymb}
\usepackage{makecell}
\usepackage{amsmath}
\usepackage{tcolorbox}
\usepackage{multirow}
\usepackage{booktabs}
\usepackage{xcolor}
\usepackage{colortbl}
\usepackage{pgfplots}
\usepackage{pgf,tikz}

\definecolor{colorsogn}{RGB}{255, 111, 97}%
\definecolor{colordsr}{RGB}{90, 91, 159}
\definecolor{colortransformer}{RGB}{240, 192, 90}
\definecolor{colortransformer2}{RGB}{60, 146, 199}
\definecolor{colorgpt3}{RGB}{217, 79, 112}
\definecolor{colorgpt4}{RGB}{0, 148, 115}

\title{Self-Organizing Agent Network for LLM-based Workflow Automation}
\author {
    Yiming Xiong\textsuperscript{\rm 1},
    Jian Wang\textsuperscript{\rm 1},
    Bing Li\textsuperscript{\rm 1},
    Yuhan Zhu\textsuperscript{\rm 1},
    Yuqi Zhao\textsuperscript{\rm 2}
}
\affiliations {
    \textsuperscript{\rm 1}Wuhan University\\
    \textsuperscript{\rm 2}Central China Normal University\\
    ymxiong@whu.edu.cn, jianwang@whu.edu.cn, bingli@whu.edu.cn, zhuyuhan2333@whu.edu.cn, yuqizhao@ccnu.edu.cn
}

\usepackage{bibentry}

\begin{document}

\maketitle

\begin{abstract}
    \input{section-abstract}
\end{abstract}


\input{section-introduction}

\input{section-related-work}

\input{section-motivation}

\input{section-methods-1}

\input{section-experiments}

\input{section-conclusion}

\appendix
\bibliography{aaai2026}

%

\input{section-appendix}

\end{document}

%% file: section-abstract.tex





Recent multi-agent frameworks built upon large language models (LLMs) have demonstrated remarkable capabilities in complex task planning. However, in real-world enterprise environments, business workflows are typically composed through modularization and reuse of numerous subprocesses, resulting in intricate workflows characterized by lengthy and deeply nested execution paths. Such complexity poses significant challenges for LLM-driven orchestration, as extended reasoning chains and state-space explosions severely impact planning effectiveness and the proper sequencing of tool invocations. Therefore, developing an orchestration method with controllable structures capable of handling multi-layer nesting becomes a critical issue. To address this, we propose a novel structure-driven orchestration framework—\textbf{Self-Organizing Agent Network (SOAN)}. SOAN incrementally builds a formalized agent network by identifying and encapsulating structural units as independent agents, enhancing modularity and clarity in orchestration. Extensive evaluations were performed using multiple benchmarks as well as a real-world enterprise workflow dataset. Experimental results demonstrate that SOAN significantly outperforms state-of-the-art methods in terms of adaptability, fault tolerance, and execution efficiency. 

%% file: section-introduction.tex
\section{Introduction}
Workflow automation is shifting from rule-based systems to LLM-driven intelligent systems \cite{DBLP:journals/corr/abs-2411-05451, DBLP:conf/icaif/ZengWCRRBV23, DBLP:conf/emnlp/XiaoMWW0WHL24}. Traditional automation focused on reducing manual labor through repetitive task execution \cite{cichocki1997workflow}. The rise of large language models (LLMs), with their capabilities in understanding natural language and generating structured content, now enables systems that not only follow instructions but also comprehend tasks, plan content, and reason through multi-step processes.

\begin{figure}[ht]
    \centering
    \includegraphics[width=\linewidth]{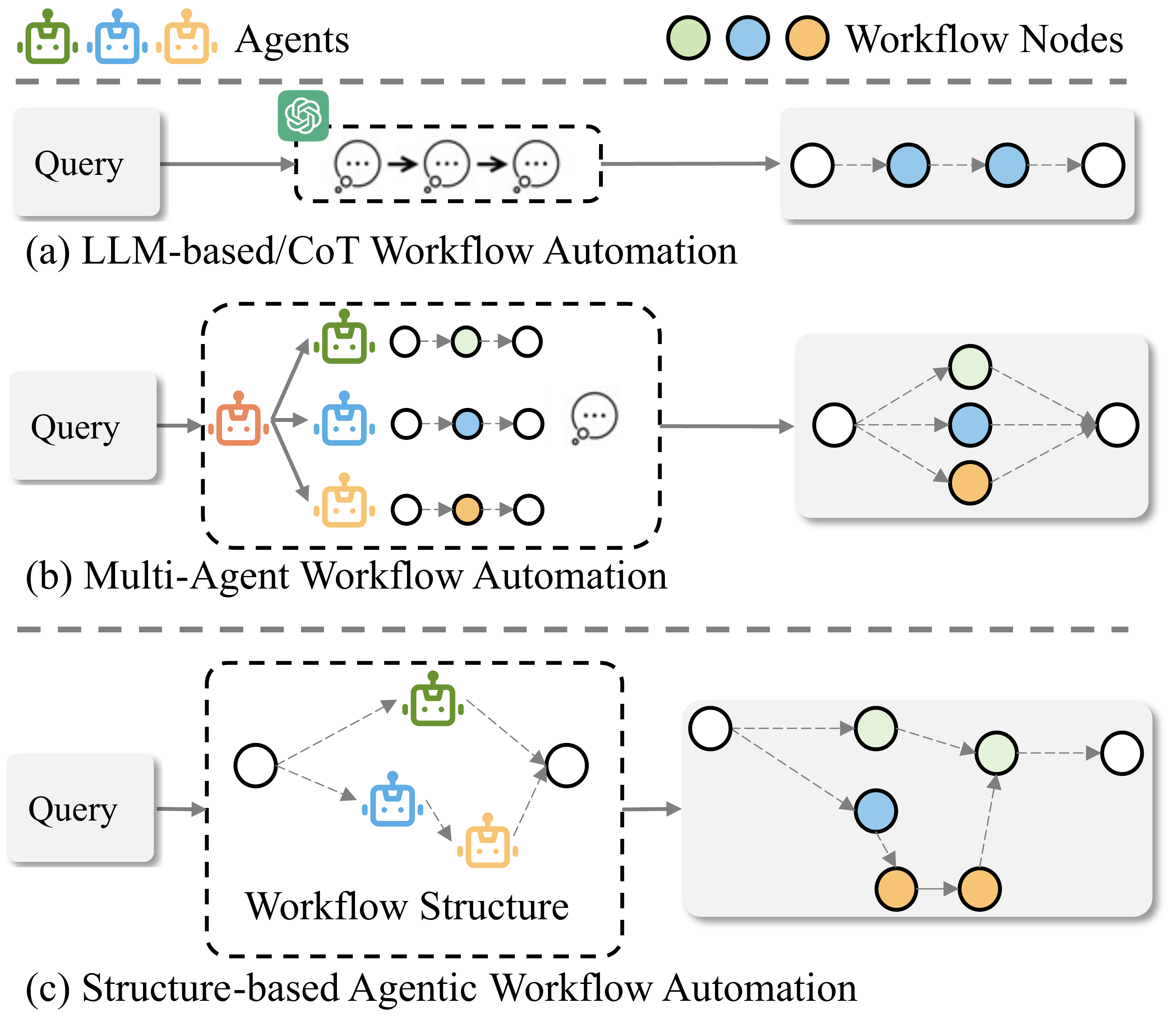}
    \caption{Different workflow automation modes based on LLMs.}
    \label{fig:Introduction}
 \end{figure}


Figure~\ref{fig:Introduction} (a) illustrates the Chain-of-Thought (CoT) mechanism \cite{DBLP:conf/nips/Wei0SBIXCLZ22, DBLP:conf/iclr/0002WSLCNCZ23} which boosts LLM-driven automation by explicitly exposing intermediate reasoning steps, thereby improving interpretability and local planning.  Nevertheless, CoT is confined to a single model’s reasoning, lacking collaborative specialization and struggling to generalize across domains.
Considering this limitation, researchers have turned to multi-agent systems (MAS). As shown in Figure~\ref{fig:Introduction} (b), LLM-based agents collaborate with each other to jointly handle complex tasks. Systems such as MetaGPT \cite{DBLP:conf/iclr/HongZCZCWZWYLZR24}, CAMEL \cite{DBLP:conf/nips/LiHIKG23}, and AutoGen \cite{DBLP:journals/corr/abs-2308-08155} use standard operating procedures (SOPs), dual-agent dialogues, and dynamic agent chains, respectively, showing good orchestration capabilities.
However, these systems face challenges in enterprise settings \cite{DBLP:conf/kbse/0010D0024, DBLP:journals/corr/abs-2501-18002}. Real-world workflows are complex in structure, often involving long sequences and nested dependencies, and are difficult to model, optimize, or reuse using current multi-agent system (MAS) approaches \cite{DBLP:conf/acl/PatelRC24}. They also require a high degree of semantic consistency and structure awareness.

To address this challenge, we propose a new framework grounded in structural principles. Enterprise workflows often exhibit repeatable patterns, modular hierarchies, and interchangeable subprocesses. SOAN effectively identifies and leverages these structural regularities to enhance agent coordination, workflow reuse, and generalization. Unlike traditional multi-agent systems, SOAN incorporates structure-aware abstractions into agent behaviors, enabling it to better manage complex, evolving enterprise processes.
Our main contributions are as follows:
\begin{itemize}
    \item We introduce a novel structure-semantic joint modeling paradigm that explicitly captures both structural patterns and semantic abstractions within workflows. This paradigm enables robust generalization to unseen workflow structures, overcoming the limitations of existing LLM-based agents that treat workflows as flat or sequential processes.

    \item We introduce SOAN (Self-Organizing Agent Network), a structure-driven multi-agent framework that captures and leverages inherent workflow patterns to guide agent collaboration. SOAN consistently outperforms existing multi-agent systems on classical reasoning and planning benchmarks, demonstrating superior generalization and robustness.

    \item We develop a structure-aware workflow design tool and build a corresponding large-scale industry dataset, gflowQA, which contains 23,520 real workflows from different industries. This dataset can be used as a rigorous benchmark to evaluate the effectiveness and robustness of SOAN in real enterprise scenarios.
    
\end{itemize}

%% file: section-related-work.tex
\section{Related Work}

\paragraph{LLM-based/CoT Workflow Automation.}
Many approaches enhance large language models (LLMs) with tool usage and step-wise reasoning to support workflow automation. ReAct~\cite{DBLP:conf/iclr/YaoZYDSN023} integrates LLMs with planning heuristics to iteratively generate tool-based action sequences. TaskMatrix.AI~\cite{liang2024taskmatrix} and ToolLLM~\cite{qin2023toolllm} further formalize tool and API interactions for real-world task execution. Chain-of-Thought prompting~\cite{DBLP:conf/nips/Wei0SBIXCLZ22, DBLP:conf/iclr/0002WSLCNCZ23} complements these by guiding LLMs through intermediate reasoning steps, enhancing accuracy and robustness.

\paragraph{Multi-Agent Workflow Automation.}
LLMs have been adopted for multi-agent collaboration in complex workflows. MetaGPT~\cite{DBLP:conf/iclr/HongZCZCWZWYLZR24} assigns structured roles (e.g., PM, engineer) to LLM agents, enabling SOP-style coordination. CAMEL~\cite{DBLP:conf/nips/LiHIKG23} enables role-driven dialogues, while AutoGen~\cite{DBLP:journals/corr/abs-2308-08155} allows dynamic agent composition and messaging. These systems demonstrate how LLM agents can support distributed planning and collective task execution~\cite{DBLP:journals/corr/abs-2503-15520, DBLP:journals/corr/abs-2408-08435, DBLP:conf/iccv/SongSWCW023, liu2024dynamic, DBLP:conf/icml/ZhugeWKFKS24, DBLP:journals/corr/abs-2407-03913, DBLP:journals/tmlr/WangX0MXZFA24, DBLP:journals/corr/abs-2502-14143}.

\paragraph{Structure-based Agentic Workflow Automation.}
Recent efforts emphasize structured representations for better modularity, planning, and interpretability. Flow~\cite{DBLP:conf/iclr/NiuSLS00L25} introduces reusable subflows; AFlow~\cite{DBLP:conf/iclr/ZhangXYTCCZCHWZ25} leverages Monte Carlo Tree Search for refinement; EvoFlow~\cite{DBLP:journals/corr/abs-2502-07373} employs heuristic search; and FlowMind~\cite{DBLP:conf/icaif/ZengWCRRBV23} incorporates user feedback and abstraction layers. These works mark a shift toward adaptive, structured orchestration~\cite{DBLP:journals/eswa/WangSH06, DBLP:journals/corr/abs-2502-05036, habler2025building, DBLP:journals/corr/abs-2502-17443, DBLP:conf/icml/ZhouYSWW24, DBLP:conf/naacl/PrasadKHCSBK24, DBLP:conf/nips/ShinnCGNY23}.

Despite these advances, current methods often struggle with coherence under structural complexity, particularly when dealing with nested tasks, evolving agent roles, or large atomic process libraries. This motivates a structure-centric agentic framework that can self-organize, dynamically optimize, and generalize across diverse workflow scenarios.

%% file: section-motivation.tex
\section{Motivation}

Real-world enterprise workflows exhibit a wide distribution in structural complexity. While many workflows consist of only one or two atomic tasks, a significant portion contains long execution paths, deep nesting, and complex control constructs.  An LLM-driven multi-agent orchestrator must therefore handle both simple and highly structured workflows with equal robustness. 

\paragraph{gflowQA dataset from enterprise environment.}

To systematically verify the generalization ability of current workflow automation methods in real scenarios, we worked with several companies to obtain 23,520 workflows in multiple cross-domain process orchestration scenarios based on unified process standards and specifications. To ensure the quality and representativeness of the dataset, we conducted rigorous data cleaning, normalization, and deduplication. After screening, we retained 8,000 high-quality workflows: 6,000 of them were used for training, and the remaining 2,000 were used for evaluation to verify their generalization ability and robustness. For details of the dataset, see Appendix~\ref{apx:gflowQA}.

\paragraph{Sensitivity of MAS to workflow depth and length.}
To assess the structural generalization capabilities of current LLM-based orchestration systems, we conducted a series of modeling and simulation experiments using three representative frameworks: AutoGen, CAMEL, and MetaGPT. These frameworks are tasked with interpreting and executing real-world workflows from our dataset, ranging from flat linear sequences to deeply nested, multi-branch structures.

\begin{figure}[h]
    \centering
    \begin{minipage}{0.45\textwidth}
        \centering

        \pgfplotstableread[]{data/result/AutoGen.txt}\DivisionSOGN
        \pgfplotstableread[]{data/result/Camel.txt}\DivisionDSR
        \pgfplotstableread[]{data/result/MetaGpt.txt}\DivisionTransformer

        \begin{tikzpicture}[scale=.8]
        \begin{axis}[
            width=4.0in,
            height=2.0in,
            xmin=1,
            xmax=8,
            xlabel=Length of workflow, 
            xtick={ 2,  3,  4, 5, 6, 7},
            ymin=-10,
            ymax=110,
            ylabel=Accuracy(\%), 
            ytick={ 0, 25, 50, 75, 100},
            axis background/.style={fill=white},
            xmajorgrids,
            ymajorgrids,
            legend style={at={(0.97,0.55)}, anchor=south east, legend cell align=left, align=left, draw=white!15!black}
        ]
            \addplot [color=colorsogn, line width=1.5pt, mark size=1.6pt, mark=*, mark options={solid, colorsogn, fill opacity=0.5}, opacity=0.6] table [x = l, y = a] from \DivisionSOGN;
            \addlegendentry{AutoGen};
            \addplot [color=colordsr, line width=1.5pt, mark size=1.6pt, mark=o, mark options={solid, colordsr, fill opacity=0.5}] table [x = l, y = a] from \DivisionDSR;
            \addlegendentry{Camel};
            \addplot [color=colortransformer, line width=1.5pt, mark size=1.6pt, mark=triangle*, mark options={solid, rotate=180, colortransformer, fill opacity=0.5}] table [x = l, y = a] from \DivisionTransformer;
            \addlegendentry{MetaGPT};
        \end{axis}
        \end{tikzpicture}
        \caption{Out-of-distribution generalization of multi-agent systems declines as workflow length increases.}
        \label{fig: task scheduling performance}
    \end{minipage}\hfill
\end{figure}
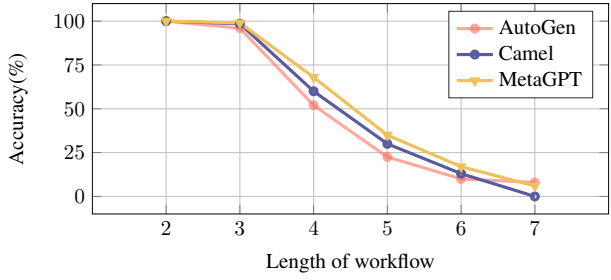

As shown in Figure~\ref{fig: task scheduling performance}, the results reveal a critical limitation: existing LLM-agent frameworks often struggle with long and complex workflows. Common failure cases include incorrect task ordering, missing dependency handling, and semantic drift during task transitions. These are not merely implementation errors—they reflect a deeper issue: the lack of explicit modeling for structural constraints and compositional logic within current LLM-agent systems.

%% file: section-methods-1.tex
\section{Structural Agentic Workflow Learning}

\begin{figure*}
    \centering
    \includegraphics[width=\textwidth]{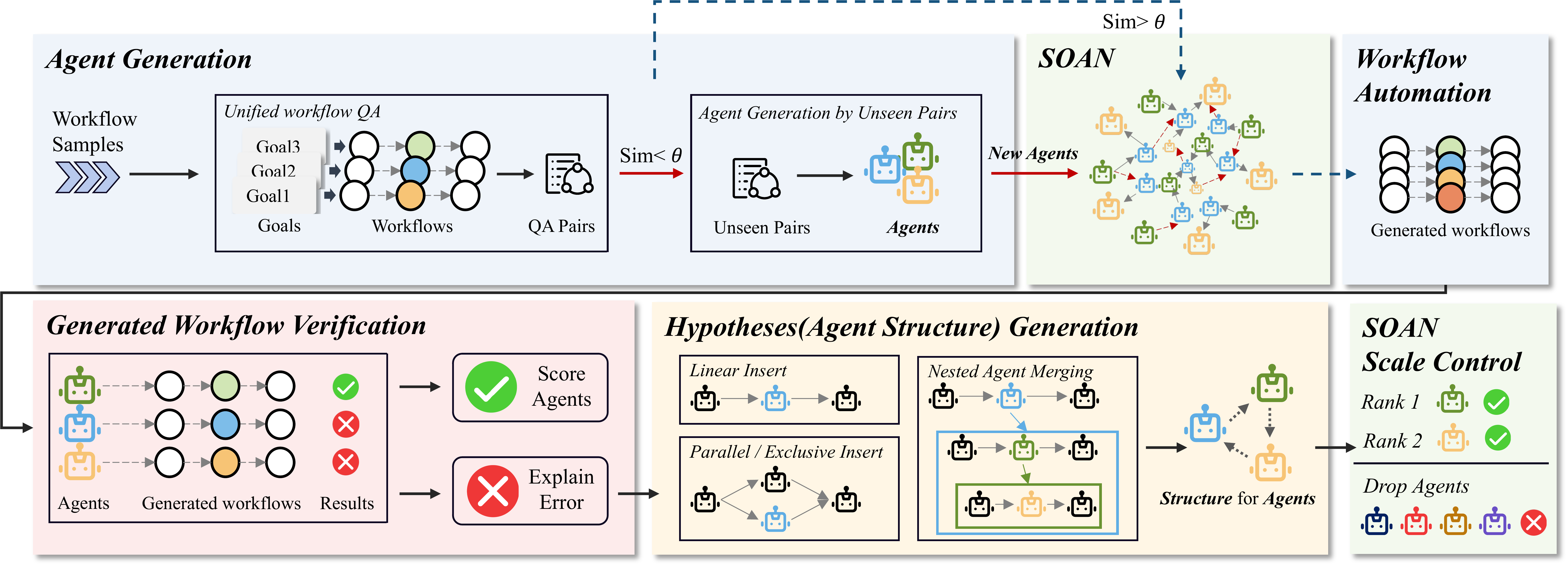}
    \caption{The overview process of our framework.}
    \label{fig:Overview}
 \end{figure*}

In this section, we present the core methodology of the proposed SOAN framework. SOAN aims to enable robust structural generalization for workflow automation, especially in scenarios where existing LLM-based agents struggle to cope with unknown or complex workflow patterns. As illustrated in Figure~\ref{fig:Overview}, the method comprises four stages:  
(1) Agent Generation (Section~\ref{sec:Agent Generation}), where specialized agents are created based on task requirements; (2) Generated Workflow Verification (Section~\ref{sec:Generated Workflow Verification}), where the produced workflow is validated for correctness; (3) Hypotheses Generation (Section~\ref{sec:Hypotheses Generation}), where optimized agent structures are synthesized to support workflow execution; (4) SOAN Scale Control (Section~\ref{sec:Scale Control}): Agents with low life values are gradually eliminated, ensuring efficient resource allocation and long-term system optimization.

\subsection{Agent Generation}\label{sec:Agent Generation}

We construct a set of \textbf{atomic agents} from numerous \emph{goal-to-atomic-procedure} pairs, where each atomic procedure is a minimal executable workflow fragment aimed at achieving a clearly defined sub-goal within a specific domain (e.g., data querying, API calls, tool triggering). These atomic workflows serve as fundamental building blocks for task execution.
Given a dataset of atomic workflows $\mathcal{D} = {(g_i, P_i)}_{i=1}^N$, where $g_i$ is a specific atomic goal (such as querying customer ID or sending email) and $P_i = \langle{t_1, t_2, \ldots, t_k}\rangle$ is the corresponding minimal workflow as an ordered set of atomic tasks $t_j$, we aim to build an agent $A_i$ that encapsulates the capability to fulfill $g_i$ by executing $P_i$.

Each workflow sample contains an execution request and a successful path. We extract five key elements for agent construction: the workflow ID (e.g., User Registration), input parameters (e.g., registration form), expected outputs (e.g., registered user), ordered execution steps, and the input-output data schemas. Formally, each atomic agent is defined as a tuple:
\begin{equation}
A_i = (g_i, P_i, \mathcal{T}_i, \mathcal{C}_i),
\end{equation}
where $g_i$ is the atomic goal the agent is designed to achieve, $P_i$ is the internal atomic process, $\mathcal{T}_i$ represents the tools or operations the agent depends on, and $\mathcal{C}_i$ denotes the context constraints or preconditions necessary for execution, such as input format requirements or domain assumptions. The main objective of an atomic agent is to learn the mapping
\begin{equation}
f : g_i \longrightarrow P_i,
\end{equation}
and given training data $\mathcal{D}$ of atomic goal–process pairs, to maximize the probability of correct execution:

\begin{equation}
\max_{A_i} \ \mathbb{P}\left( A_i(g_i) = P_i \mid (g_i, P_i) \in \mathcal{D} \right).
\end{equation}

Through this process, each atomic workflow is encapsulated as an agent annotated with its functional scope, toolset, and execution context, forming the modular and reusable foundation of the SOAN framework.

\subsection{Generated Workflow Verification}\label{sec:Generated Workflow Verification}

When encountering similar but previously unseen goals, SOAN leverages agent collaboration to dynamically construct new workflows. Instead of rigidly reusing predefined templates, the framework identifies a set of relevant agents whose capabilities align with the task objective. These agents autonomously collaborate by proposing partial solutions based on their respective toolchains and domain expertise.

\begin{figure}[h]
    \centering
    \includegraphics[width=\linewidth]{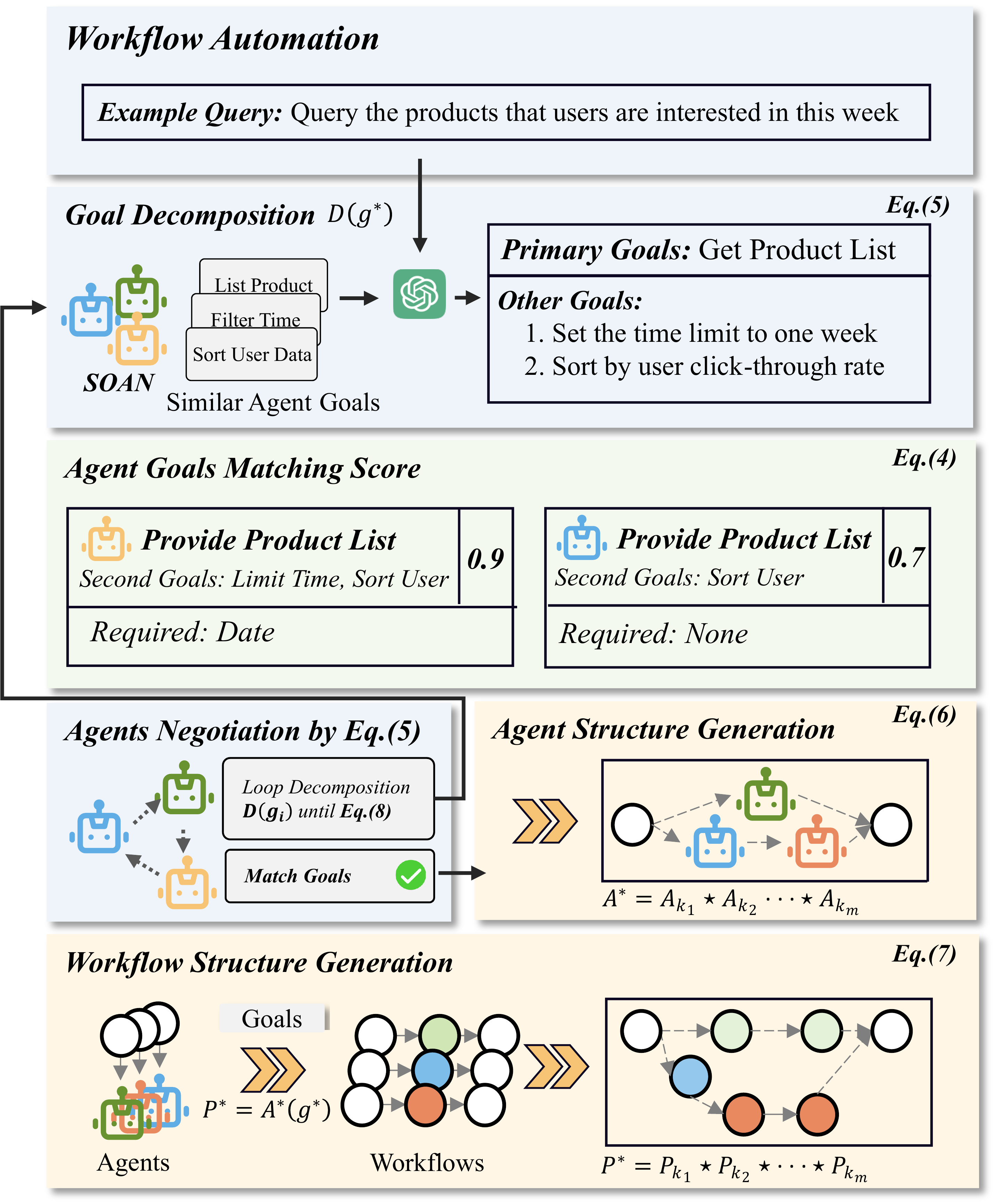}

    \caption{SOAN's Running Progress.}
    \label{fig:Running Progress} 
 \end{figure}

As illustrated in Figure~\ref{fig:Running Progress}, agents iteratively communicate and reason to compose candidate workflows by integrating atomic procedures into more complex structures. This collaborative mechanism allows SOAN to generalize beyond previously observed workflows, enabling adaptation to novel goals by reconfiguring and recombining known structural patterns. Such dynamic self-organization makes SOAN particularly well-suited for enterprise workflow environments, where goals are often diverse, evolving, and context-dependent.

Suppose the current task goal is $g^{\ast}$, which is not completely equivalent to any $g_i \in \mathcal{G}_k$ in the existing agent set ${A_k}$, but through semantic similarity or task intent mapping, a set of related agents $\mathcal{A}^* = {A_{k_1}, A_{k_2}, ..., A_{k_m}}$ can be retrieved, satisfying:
\begin{equation}
\exists A_{k} \in \mathcal{A}^*, \ \text{Sim}(g^{\ast}, \mathcal{G}_k) > \theta,
\end{equation}
where $\text{Sim}(g^{\ast}, \mathcal{G}_k)$ represents the similarity based on semantic, domain, or historical task mapping, and $\theta$ is an adjustable threshold. 

Let the goal decomposition function be \( D(g^*) \), which recursively decomposes a complex goal into a set of subgoals as follows:

\begin{equation}
D(g^*) = 
\begin{cases}
A_k(g_k), & \exists A_k \in \mathcal{A}^*, \text{Sim}(g^*, \mathcal{G}_k) > \theta \\
\displaystyle \bigcup_{i=1}^n D(g_i), &  g^* = \{g_1, g_2, \ldots, g_n\}
\end{cases}
\end{equation}
If there exists a sub-agent \( A_k \) that can directly handle the goal \( g^* \) (i.e., similarity exceeds the threshold), the decomposition stops and the goal itself is returned;
Otherwise, the goal \( g^* \) is decomposed into a set of subgoals \( \{g_1, g_2, \ldots, g_n\} \), and \( D \) is recursively called on each subgoal.

Through the collaboration of multiple agents in $\mathcal{A}^*$, a candidate process $P^{\ast}$ that can cover the requirements of $g^{\ast}$ is combined:

\begin{align}
   A^{\ast}(g^\ast) = A_{k_1}(g_{k_1}) \star \dots \star A_{k_m}(g_{k_m}),\\
    P^{\ast} = A^{\ast}(g^\ast) = P_{k_1} \star P_{k_2} \star \dots \star P_{k_m},
\end{align}
where $P_{k_i}$ represents the atomic process maintained inside $A_{k_i}$, and $\star$ represents the structural splicing operation (concatenation, branching, nesting).
The final collaboration process should meet the following requirements:

\begin{equation}
    \mathbb{P}\left( P^{\ast} \rightarrow g^{\ast} \right) > \eta,
\end{equation}
where $\eta$ is the preset reliability threshold of the system (usually close to 1).
By comparing the generated $P^{\ast}$ with the expected workflow, we can get the correctness of the current Agent execution result.

\subsection{Hypotheses Generation}\label{sec:Hypotheses Generation}



Through agent collaboration, the system can construct a candidate workflow $P^{\ast}$ for a novel goal $g^{\ast}$ based on previously learned atomic procedures. However, since $g^{\ast}$ is not present in the training data, the generated workflow $P^{\ast}$ is not guaranteed to be correct. To address this, SOAN introduces a feedback-driven structural optimization mechanism that enables hypothesis revision and iterative refinement of the workflow.

For successful workflows, agents receive positive reinforcement, and their structural patterns are preserved as reusable knowledge. For failed workflows, SOAN generates structural hypotheses to explain the failure and applies targeted structural refinements through three core operations:\\
\textbf{Linear Insertion.}
To address missing preconditions or incomplete toolchains, SOAN performs linear insertion by integrating an auxiliary agent $A_{\text{new}}$ into an existing sequence $A$. This operation is denoted as:
\begin{equation}
    A \star A_{\text{new}} = A \oplus A_{\text{new}},
\end{equation}
where $\oplus$ represents the structural augmentation operator that merges $A_{\text{new}}$ into the execution path of $A$. The inserted agent fills in execution gaps or resolves unmet dependencies, ensuring that the workflow remains functionally complete and structurally coherent.\\
\textbf{Branching.}
To handle divergent scenarios and increase fault tolerance, SOAN introduces conditional branching by associating specific agents with triggering conditions. Formally:
\begin{equation}
    A \star A_{\text{new}} = A \oplus \{\text{Cond}  \rightarrow A_{\text{new}}\},
\end{equation}
this enables the workflow to flexibly adapt to edge cases, exception handling, or user-specific configurations without modifying the main execution path.\\
\textbf{Nesting.}
To support modularity and composability, SOAN nests agents by recursively decomposing their associated goals into finer-grained subgoals. Given two agents $A(g)$ and $A_{\text{new}}(g_{\text{new}})$, nesting is defined as:
\begin{align}
    A(g) \star A_{\text{new}}(g_{\text{new}}) = A \star (D(g) \cup D(g_{\text{new}})),
\end{align}
where $D(\cdot)$ is the recursive decomposition function. This operation replaces abstract goals with their decomposed substructures, allowing SOAN to encapsulate complex logic within hierarchical agent modules. Such nested forms enable better structural generalization, reuse of common subflows, and clearer abstraction boundaries in large-scale workflows.

\subsection{Network Scale Control}\label{sec:Scale Control}
In SOAN, each agent is treated as an adaptive execution entity endowed with a dynamic life-value that reflects its historical utility, stability, and generalization performance. This life-based agent control mechanism allows for evolutionary orchestration strategies, enabling the system to retain structurally adaptive agents and eliminate underperforming ones.


Each agent $A_i$ is assigned a scalar life value $L_i \in [0, L_{\text{max}}]$. Initially, all agents are initialized with $L_i = L_{\text{init}}$. Over time, $L_i$ is updated based on the agent’s task execution behavior:

\begin{equation}
   L_i \leftarrow L_i + \Delta L^{+}_i - \Delta L^{-}_i,
\end{equation}
where $\Delta L^{+}_i$ reward gain based on success, generalization, and reuse; $\Delta L^{-}_i$ penalty from execution failure, semantic drift, or redundancy.


We define reward and penalty components as follows:
\paragraph{Reward Agents ($\Delta L^{+}_i$).}
$R_c$ rewards correct execution where the agent output matches ground truth. $R_s$ rewards symmetry-based reuse across structurally equivalent tasks. $R_g$ rewards successful generalization to unseen tasks through abstraction.
\paragraph{Penalty Agents ($\Delta L^{-}_i$).}
$P_e$ penalizes task failures. $P_s$ penalizes semantic drift from expected intent. $P_r$ penalizes redundant outputs with no structural gain.

Formally:
\begin{align}
\Delta L^{+}_i &= \alpha_1 R_c + \alpha_2 R_s + \alpha_3 R_g, \\
\Delta L^{-}_i &= \beta_1 P_e + \beta_2 P_s + \beta_3 P_r,
\end{align}
where $\alpha_k$, $\beta_k$ are tunable per task domain.During Agent Net execution, when multiple agents qualify for a transition $t_j$, their selection probability is computed as:
\begin{equation}
\mathbb{P}(A_i \mid t_j) = \frac{L_i \cdot \gamma_i(t_j)}{\sum_{k} L_k \cdot \gamma_k(t_j)},
\end{equation}
where $\gamma_i(t_j)$ is a compatibility score based on workflow match (hard constraint), structural familiarity, and prior execution similarity.

\paragraph{Agent Elimination and Replacement}
Agents with $L_i \leq 0$ are moved to an archive. The active pool $\mathcal{A}$ is periodically refreshed from archived or newly spawned agents.

%% file: section-experiments.tex
\section{Experiments}
We evaluate the effectiveness of SOAN through four key research questions on structurally diverse workflow benchmarks and real-world cross-domain datasets: 
\begin{itemize}
    \item{RQ1:} Can SOAN outperform existing agent orchestration frameworks in terms of structural generalization accuracy, agent reuse rate, and workflow efficiency?
    \item{RQ2:} Can SOAN outperform existing LLM-based multi-step inference methods?
    \item{RQ3:} Can SOAN effectively scale with the increasing number of atomic processes in enterprise workflows?
    \item{RQ4:} Can the structure-guided optimization mechanism improve SOAN’s accuracy in workflow generation?
\end{itemize}

\input{section-RQ-agent-compare.tex}

\input{section-RQ-accuracy.tex}

\input{section-RQ-atomic-flow}

\input{section-RQ-ablation-sym.tex}

%% file: section-RQ-agent-compare.tex
\subsection{RQ1: Comparison in Workflow Orchestration}

\paragraph{Dataset and Setting.} To evaluate the generalization capability of different agent-based workflow systems under varying structural complexity, we constructed a benchmark gflowQA (Appendix~\ref{apx:gflowQA}) containing task scheduling workflows with distinct control structures and node sizes. The benchmark is divided into two categories based on structural type:
\begin{itemize}
    \item{Linear workflows:} exhibit length-based complexity, with tasks arranged in flat, sequential chains. As the number of steps increases (e.g., 2–3 to 7+), models must maintain consistency over longer execution paths, testing their ability to handle extended procedural dependencies.

    \item{Nested workflows:} reflect depth-based complexity, where tasks are achieved via recursive sub-workflow calls. As nesting deepens (e.g., 1–2 to 5+ layers), models must reason across abstraction layers, resolve inter-module dependencies, and manage hierarchical execution coherence.

\end{itemize}
Each category is further subdivided based on the number of task nodes: (1) Small (2–3 for linear, 1–2 for nested), (2) Medium (4–6 for linear, 3–4 for nested), (3) Large (7+ for linear, 5+ for nested).This allows for a more detailed observation of changes in accuracy.
The goal is to correctly schedule and complete the workflow, and the systems are evaluated using pass@k metrics (k = 1, 3, 5), which measure the proportion of correct task plans within the top-k generated candidates.
\paragraph{Comparison Methods and Results.}
We select four state-of-the-art agent-based workflow schedulers for comparison: AutoGen\cite{DBLP:journals/corr/abs-2308-08155}, a coordination-first framework utilizing LLMs as backend agents;
Camel\cite{DBLP:conf/nips/LiHIKG23}, which applies role-based prompting and reflection for cooperative agent planning; MetaGPT\cite{DBLP:conf/iclr/HongZCZCWZWYLZR24}, a code-oriented multi-agent system with modular role templates and stepwise collaboration; 
All methods use the same model backend (GPT-4o) for fairness.

Table~\ref{tab:cross-dataset-performance} presents the task scheduling accuracy across different workflow complexities. The results reveal several key insights: In simple linear workflows (2–3 nodes), all methods achieve high pass@1 scores (above 91\%), with SOAN slightly outperforming others (95.1\%).

\input{section-result-agent.tex}

As the workflow depth increases, the performance gap widens. In linear workflows with 7+ nodes, SOAN maintains a high pass@1 of 91.6\%, while others drop sharply. The difference is more pronounced in nested workflows, where AutoGen and MetaGPT degrade to 37.2\% and 34.9\% pass@1 respectively on deeply nested flows (5+ nodes). SOAN, however, sustains a robust 89.3\% pass@1, reflecting its superior handling of control structures and recursion.

Even in moderately nested cases (3–4 nodes), SOAN consistently leads by over 30 points compared to baseline systems.
These results validate SOAN’s ability to generalize structures, especially in the presence of increased nesting and depth, which traditional LLM-based multi-agent planners struggle with due to their lack of global coordination or structure awareness. The significant performance gains highlight the benefits of incorporating symmetry-guided structural optimization into workflow generation.

%% file: section-result-agent.tex
\begin{table}[ht]
    \centering
    \small
    \begin{tabular}{cccccc}
    \toprule
    \textbf{Type} & \textbf{Nodes} & \textbf{Method} & \makecell[c]{\textbf{pass@1}}  & \makecell[c]{\textbf{pass@3}} & \makecell[c]{\textbf{pass@5}} \\
    \midrule
    \multirow{12}{*}{Linear}
    & \multirow{4}{*}{2--3} 
        & AutoGen & 91.3 & 96.0 & 97.5 \\
        &         & Camel   & 94.7 & 97.2 & 98.1 \\
        &         & MetaGPT & 93.5 & 96.5 & 97.6 \\
        &         & SOAN    & 95.1 & 97.8 & 98.3 \\
    \cline{2-6}
    & \multirow{4}{*}{4--6} 
        & AutoGen & 82.6 & 89.4 & 91.1 \\
        &         & Camel   & 86.2 & 92.3 & 93.7 \\
        &         & MetaGPT & 84.5 & 90.2 & 92.0 \\
        &         & SOAN    & 93.8 & 97.5 & 98.1 \\
    \cline{2-6}
    & \multirow{4}{*}{7+} 
        & AutoGen & 64.3 & 76.2 & 80.0 \\
        &         & Camel   & 70.4 & 82.9 & 86.2 \\
        &         & MetaGPT & 61.0 & 75.3 & 79.1 \\
        &         & SOAN    & 91.6 & 96.9 & 97.8 \\
    \hline
    \multirow{12}{*}{Nested}
    & \multirow{4}{*}{1--2} 
        & AutoGen & 69.8 & 81.2 & 84.3 \\
        &         & Camel   & 76.1 & 87.3 & 89.7 \\
        &         & MetaGPT & 71.5 & 84.0 & 86.1 \\
        &         & SOAN    & 92.2 & 96.5 & 97.3 \\
    \cline{2-6}
    & \multirow{4}{*}{3--4} 
        & AutoGen & 54.0 & 70.3 & 74.6 \\
        &         & Camel   & 60.8 & 76.2 & 79.5 \\
        &         & MetaGPT & 52.1 & 68.0 & 72.3 \\
        &         & SOAN    & 90.7 & 95.9 & 97.0 \\
    \cline{2-6}
    & \multirow{4}{*}{5+} 
        & AutoGen & 37.2 & 58.1 & 63.4 \\
        &         & Camel   & 45.6 & 66.3 & 71.0 \\
        &         & MetaGPT & 34.9 & 53.6 & 59.7 \\
        &         & SOAN    & 89.3 & 95.2 & 96.6 \\
    \bottomrule 
    \end{tabular}
    \caption{Complex workflows orchestrating performance of AutoGen, Camel, and MetaGPT under different workflow structures.}
    \label{tab:cross-dataset-performance}
\end{table}

%% file: section-RQ-accuracy.tex
\subsection{RQ2: Comparison in Multi-step Inference}
This set of experiments aims to evaluate the generalization, reasoning accuracy, and structure-aware learning capability of our proposed system, SOAN (Self-Organizing Agent Network), across diverse domains including textual inference (T-Eval), ontology-based question answering (PrOntoQA), reasoning over unseen ontology schema (PrOntoQA-OOD), multi-hop QA (HotpotQA), math reasoning (GSM8K), and structured workflow reasoning (gFlowQA). All datasets are split into training and held-out test sets. Each input instance is given in symbolic form or natural language format depending on the task domain. The underlying model for all methods is GPT-4o, and we apply consistent inference protocols (e.g., 5-shot prompting for CoT-SC, multiple rounds for Self-Refine and SOAN) across all methods for fair comparison.

Our proposed SOAN framework incorporates a modular reasoning structure. Each workflow is dynamically generated and refined during the inference phase based on feedback, leveraging task-specific knowledge embeddings and self-organizing control modules. For each benchmark, we measure performance using accuracy (exact match or final numerical correctness), averaged over three randomized runs.

\input{section-result-on-gFlow.tex}

\paragraph{Compared Methods and Results.}
We compare our method with a suite of established baselines:
(1) IO (GPT-4o): vanilla GPT-4o using direct input-output prompting;
(2) Chain-of-Thought\cite{DBLP:conf/nips/Wei0SBIXCLZ22}, which augments input with intermediate reasoning;
(3) CoT-SC\cite{DBLP:conf/iclr/0002WSLCNCZ23}, which applies few-shot self-consistency decoding;
(4) MultiPersona\cite{DBLP:conf/naacl/WangMW0WJ24}, which prompts with multiple specialized agent personas;
(5) Self-Refine\cite{DBLP:conf/nips/MadaanTGHGW0DPY23}, an iterative refinement framework;
(6) AFlow\cite{DBLP:conf/iclr/ZhangXYTCCZCHWZ25}, an agent workflow planning system optimized by reinforcement learning.

The performance is reported in Table~\ref{tab:qa-code-math-comparison}, where we see that SOAN consistently outperforms all baselines across tasks. Specifically, SOAN achieves 88.3\% on T-Eval, 99.5\% on PrOntoQA, and a substantial 90.8\% on PrOntoQA-OOD, showing strong structural generalization. On more complex multi-step reasoning tasks like HotpotQA and gFlowQA, SOAN surpasses the best baseline (AFlow) by over 10\% absolute accuracy. 



%% file: section-result-on-gFlow.tex
\begin{table*}[ht]
    \centering
    \begin{tabular}{l|ccccc|c}
    \toprule
    \textbf{Method} & \textbf{T-Eval} & \textbf{PrOntoQA} & \textbf{PrOntoQA-OOD} & \textbf{HotpotQA} & \textbf{GSM8K} & \textbf{gflowQA} \\
    \midrule
    IO (GPT-4o) & 86.4 & 99.6 & 83.0 & 70.3 & 92.7 & 48.6 \\
    \midrule
    CoT\cite{DBLP:conf/nips/Wei0SBIXCLZ22} & 87.5 & 98.5 & 74.3 & 71.4 & 92.5 & 50.2 \\
    CoT-SC (5-shot)\cite{DBLP:conf/iclr/0002WSLCNCZ23} & 85.1 & 96.1 & 79.0 & 73.5 & 92.5 & 53.0 \\
    MultiPersona\cite{DBLP:conf/naacl/WangMW0WJ24} & 84.3 & 94.4 & 65.8 & 73.6 & 92.8 & 52.7 \\
    Self-Refine\cite{DBLP:conf/nips/MadaanTGHGW0DPY23} & 80.6 & 79.8 & 70.2 & 57.3 & 87.5 & 46.1 \\
    \midrule
    AFlow\cite{DBLP:conf/iclr/ZhangXYTCCZCHWZ25} & 86.2 & 96.1 & 81.3 & 74.2 & 90.5 & 61.2 \\
    \cellcolor{gray!25}\textbf{SOAN (Ours)} & \cellcolor{gray!25}\textbf{88.3} & \cellcolor{gray!25}\textbf{99.5} & \cellcolor{gray!25}\textbf{90.8} & \cellcolor{gray!25}\textbf{76.8} & \cellcolor{gray!25}\textbf{92.4} & \cellcolor{gray!25}\textbf{90.1} \\
    \bottomrule
    \end{tabular}
    
    \caption{Performance comparison between manually designed methods and workflows generated by automated workflow optimization in QA, code, and math scenarios. All methods are executed with GPT-4o on a divided test set, and results are averaged over three runs.}
    \label{tab:qa-code-math-comparison}
\end{table*}

\begin{table*}[ht]
    \centering
    \begin{tabular}{l|cccccc}
        \toprule
        \textbf{Model Variant} & \textbf{T-Eval} & \textbf{PrOntoQA} & \textbf{PrOntoQA-OOD} & \textbf{HotpotQA} & \textbf{GSM8K} & \textbf{gFlowQA} \\
        \midrule
        - No Scale Control & 76.5 & 85.3 & 81.0 & 72.1 & 81.5 & 72.4 \\
        - No Workflow Verification & 45.4 & 54.2 & 52.0 & 44.3 & 59.8 & 41.6 \\
        - No Structure Hypothesis & 0 & 0 & 0 & 0 & 0 & 0 \\
        \midrule
        Full SOAN & \textbf{88.3} & \textbf{99.5} & \textbf{90.8} & \textbf{76.8} & \textbf{92.4} & \textbf{90.1} \\
        \bottomrule
    \end{tabular}
    \caption{Ablation study of SOAN components across six benchmark datasets. Removing each module causes a noticeable drop in performance, especially on structurally complex tasks (e.g., PrOntoQA-OOD and gFlowQA).}
    \label{tab:ablation}
\end{table*}

%% file: section-RQ-atomic-flow.tex
\subsection{RQ3: Comparison in Scalability} 

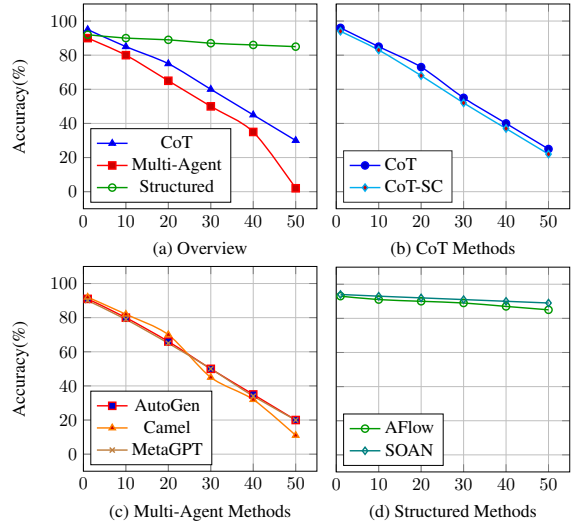
\begin{figure}[h]
    \centering
    \begin{tikzpicture}[scale=.68] 

        \begin{axis}[
            width=2.42in,
            height=2.2in,
            xmin=0,
            xmax=55,
            xlabel=(a) Overview, 
            xtick={0,10,20,30,40,50},
            ymin=-10,
            ymax=110,
            ylabel=Accuracy(\%), 
            ytick={ 0, 20, 40, 60, 80, 100},
            axis background/.style={fill=white},
            xmajorgrids,
            ymajorgrids,
            legend pos=south west,
        ]
            \addplot+[smooth, thick, blue, mark=triangle*]  coordinates {(1,95)(10,85)(20,75)(30,60)(40,45)(50,30)};
            \addlegendentry{CoT}
    
            \addplot+[smooth, thick, red, mark=square*]  coordinates {(1,90)(10,80)(20,65)(30,50)(40,35)(50,2)};
            \addlegendentry{Multi-Agent}
    
            \addplot+[smooth, thick, green!60!black, mark=o]  coordinates {(1,92)(10,90)(20,89)(30,87)(40,86)(50,85)};
            \addlegendentry{Structured}
        \end{axis}
    \end{tikzpicture}
    \begin{tikzpicture}[scale=.68]

        \begin{axis}[
            width=2.42in,
            height=2.2in,
            xmin=0,
            xmax=55,
            xlabel=(b) CoT Methods, 
            xtick={0,10,20,30,40,50},
            yticklabels=\empty,
            ymin=-10,
            ymax=110,
            ytick={ 0, 20, 40, 60, 80, 100},
            axis background/.style={fill=white},
            xmajorgrids,
            ymajorgrids,
            legend style={at={(0.5,0.425)}, anchor=south east, legend cell align=left, align=left, 
            legend pos=south west,
            draw=white!15!black}
        ]
            \addplot+[smooth, thick, blue, mark=*]  coordinates {(1,96)(10,85)(20,73)(30,55)(40,40)(50,25)};
            \addlegendentry{CoT}
        
            \addplot+[smooth, thick, cyan, mark=diamond*]  coordinates {(1,94)(10,83)(20,68)(30,52)(40,37)(50,22)};
            \addlegendentry{CoT-SC}
        \end{axis}
    \end{tikzpicture}
    \begin{tikzpicture}[scale=.68]

        \begin{axis}[
            width=2.42in,
            height=2.2in,
            xmin=0,
            xmax=55,
            xlabel=(c) Multi-Agent Methods, 
            xtick={0,10,20,30,40,50},
            ymin=-10,
            ymax=110,
            ylabel=Accuracy(\%), 
            ytick={ 0, 20, 40, 60, 80, 100},
            axis background/.style={fill=white},
            xmajorgrids,
            ymajorgrids,
            legend pos=south west,
        ]
            \addplot+[smooth, thick, red, mark=square*]  coordinates {(1,91)(10,80)(20,66)(30,50)(40,35)(50,20)};
            \addlegendentry{AutoGen}
    
            \addplot+[smooth, thick, orange, mark=triangle*]  coordinates {(1,92)(10,82)(20,70)(30,45)(40,32)(50,11)};
            \addlegendentry{Camel}
    
            \addplot+[smooth, thick, brown, mark=x]  coordinates {(1,90)(10,79)(20,65)(30,50)(40,34)(50,20)};
            \addlegendentry{MetaGPT}
        \end{axis}
    \end{tikzpicture}
    \begin{tikzpicture}[scale=.68]

        \begin{axis}[
            width=2.42in,
            height=2.2in,
            xmin=0,
            xmax=55,
            xlabel=(d) Structured Methods, 
            xtick={0,10,20,30,40,50},
            yticklabels=\empty,
            ymin=-10,
            ymax=110,
            ytick={ 0, 20, 40, 60, 80, 100},
            axis background/.style={fill=white},
            xmajorgrids,
            ymajorgrids,
            legend pos=south west
        ]
            \addplot+[smooth, thick, green!60!black, mark=o]  coordinates {(1,93)(10,91)(20,90)(30,89)(40,87)(50,85)};
            \addlegendentry{AFlow}
    
            \addplot+[smooth, thick, teal, mark=diamond]  coordinates {(1,94)(10,93)(20,92)(30,91)(40,90)(50,89)};
            \addlegendentry{SOAN}
        \end{axis}
    \end{tikzpicture}

    \caption{Accuracy decreases as the number of atomic workflows increases across different workflow automation paradigms.}
    \label{fig:Accuracy Results}
\end{figure}

We conduct a comprehensive evaluation of the three major paradigms—Chain-of-Thought (CoT), Multi-Agent, and Structured workflows—on workflows of increasing tool complexity (i.e., the number of atomic operations per process). Figure~\ref{fig:Accuracy Results} presents four subfigures, each corresponding to a comparison scope.

The overall trend shows that as tool complexity increases, CoT-based and Multi-Agent frameworks degrade significantly. CoT accuracy drops from over 90\% to below 40\% beyond 30 atomic workflows, while Multi-Agent methods (AutoGen and Camel) suffer even sharper declines due to coordination overhead and weak structural modeling. In contrast, Structured frameworks (AFlow and SOAN) maintain over 85\% accuracy with 40+ atomic workflows, demonstrating greater robustness in complex planning.

\textbf{CoT.} Comparing vanilla CoT and structure-conditioned CoT (Cot-SC) with GPT-4o, Cot-SC performs slightly better in mid-complexity, but both deteriorate beyond 25 atomic workflows, revealing limitations of linear reasoning under high inter-tool dependencies.

\textbf{Multi-Agent.} AutoGen, Camel, and MetaGPT struggle under the same setup. Camel leverages role prompting, AutoGen uses delegation, yet none sustain above 50\% accuracy with 35+ steps, constrained by coordination bottlenecks and lack of explicit structure.

\textbf{Structured Methods.} SOAN outperforms AFlow in high-complexity tasks, leveraging symmetry-aware planning to enable accurate subgoal decomposition and path recovery, maintaining over 85\% accuracy with 50 atomic workflows, validating the strength of structural priors and optimization.

%% file: section-RQ-ablation-sym.tex
\subsection{RQ4: Ablation Analysis} 

Table~\ref{tab:ablation} presents ablation results evaluating each core module's contribution in SOAN across six benchmarks. The analysis shows that removing certain modules causes significant performance drops, especially on structurally complex tasks like PrOntoQA-OOD and gFlowQA. Removing the Scale Control module, which dynamically adapts agent orchestration to workflow depth and length, leads to moderate accuracy declines on all datasets, with the largest drop on gFlowQA, highlighting the importance of scale management for long and nested workflows. Disabling the Workflow Verification module, responsible for validating structural and semantic consistency during intermediate planning, causes severe accuracy losses exceeding 30\% on most tasks, demonstrating its vital role in preserving execution integrity and preventing cascading failures in multi-step reasoning. Most notably, eliminating the Structure Hypothesis component that aligns agents and task structures results in total failure with zero accuracy across all benchmarks, indicating that structure-based reasoning is fundamental to SOAN’s agent-task matching and compositional generalization.

%% file: section-conclusion.tex
\section{Conclusion}

This paper presents SOAN, a self-organizing agent network designed to enhance workflow automation through advanced structural modeling and optimization. SOAN combines agent collaboration with structural generalization, enabling dynamic adaptation to unseen workflows by leveraging structural hypotheses. Experimental results show that SOAN consistently outperforms existing LLM-based and multi-agent baselines in workflow generalization tasks. Notably, SOAN excels at handling nested workflows, novel goals, and complex control flows with improved robustness and accuracy. This highlights the importance of structural priors in enabling scalable, interpretable, and reusable orchestration for workflow automation.

In the future, we plan to enhance SOAN with more powerful transaction control mechanisms to ensure consistency and fault recovery in multi-step workflows. We will also improve the robustness of multi-agent coordination under partial failures, and noisy inputs. 


%% file: section-appendix.tex
\newpage

\input{section-threats.tex}

\input{section-appendix-gflowQA}

\input{section-appendix-supplementary-experiments}

%% file: section-threats.tex
\section{Discussion}

\subsection{Limitation}
While SOAN demonstrates promising results in workflow structure generalization and agent orchestration, several limitations should be noted.
\textbf{1.} SOAN assumes the availability of reusable atomic workflows and stable structural patterns, which are common in enterprise domains but less applicable in open-ended or highly creative domains where workflows are ad hoc or rapidly evolving. This assumption may limit SOAN’s generalizability beyond structured business processes.
\textbf{2.} Current structural optimization approaches primarily emphasize transformations such as insertion, branching, and nesting, but often overlook deeper semantic reasoning about domain-specific constraints and interdependencies. Consequently, SOAN may generate workflows that are structurally valid yet semantically sub-optimal in highly specialized domains.

Future work will explore more adaptive optimization strategies, robustness against noisy feedback, and dynamic agent evolution to extend SOAN’s applicability to broader and more dynamic environments.

\subsection{Threats to Validity}
While SOAN demonstrates strong capabilities in workflow structure generalization and dynamic agent orchestration, several potential threats to validity should be acknowledged regarding the method itself.

\textbf{Internal Validity.}
Our method relies on the decomposition of tasks into atomic goals and the explicit modeling of agent collaboration through structural composition. Although this aligns with common enterprise workflow design practices, the correctness of atomic goal definitions and agent capability modeling heavily depends on manual curation or domain heuristics. Errors or omissions at this stage may propagate through the system, affecting overall performance. Furthermore, the success of structure hypothesis repair and group-theoretic transformations relies on the quality of initial feedback signals, which could be noisy or incomplete in practice.

\textbf{External Validity.}
SOAN assumes the existence of reusable atomic processes and structural regularities across workflows. While this is empirically true in many enterprise domains, there may exist highly dynamic or creative domains (e.g., research workflows, artistic creation) where such structural reuse is sparse or inapplicable. Thus, SOAN's generalizability to open-ended, weakly structured tasks remains unverified.

\textbf{Construct Validity.}
We design SOAN to improve structural generalization, agent coordination accuracy, and workflow completion success through explicit structural modeling. However, these constructs emphasize correctness and compositionality over other industrial concerns such as human interpretability, cost-efficiency, or regulatory constraints. Therefore, SOAN’s current evaluation does not cover all dimensions of practical deployment utility.

\textbf{Conclusion Validity.}
Although our results consistently show SOAN outperforming existing LLM-based and multi-agent baselines, these comparisons are bounded by our specific experimental settings, agent definitions, and benchmark construction. Variations in agent design, domain distribution shifts, or alternative orchestration strategies may lead to different outcomes.

Future work will further investigate SOAN's robustness across more diverse agent ecosystems, noisier environments, and domains beyond structured enterprise workflows.

%% file: section-appendix-gflowQA.tex
\section{Detail of gflowQA Dataset} \label{apx:gflowQA}
All workflows have been normalized into a unified process specification, which includes standardized metadata, explicit input/output definitions, and a formalized node structure. This normalization ensures consistency across domains and facilitates systematic structural analysis. Specifically, the dataset has the following features: (1) The number of nodes per workflow ranges from 1 to 20, covering both simple and highly complex processes. (2) Each workflow has well-defined input/output fields, enabling accurate interface modeling and downstream validation. (3) Process nesting is explicitly modeled, supporting hierarchical composition, path compression, and symmetry-based structure learning. This benchmark not only reflects real-world industry constraints, but also provides a challenging and diverse testbed for evaluating SOAN's ability to generalize to unknown structures, adapt to heterogeneous targets, and autonomously evolve process compositions through structural optimization.


The benchmark focuses on three core reasoning abilities:

\begin{enumerate}
    \item \textbf{Multi-step Planning}: Inferring correct atomic action ordering to fulfill business intent.
    \item \textbf{Nested Logic Inference}: Identifying when and how subflows (e.g., conditionals) are required.
    \item \textbf{Structural Generalization}: Adapting to longer and deeper workflows beyond training scope.
\end{enumerate}

gflowQA is designed to test whether models such as SOAN and LLM-based baselines can scale from simple flat sequences to deeply nested, multi-step decision workflows.

\subsection{Multi-Node Workflow Distribution}
We analyze 23,521 enterprise workflows to understand the distribution of atomic operation nodes within each workflow. Among these, 9,013 workflows contain more than one atomic operation node, indicating a high proportion of multi-step, non-trivial tasks.

\begin{figure}[ht]
    \centering
    \begin{tikzpicture}
        \begin{axis}[
            width=0.9\linewidth,
            height=6cm,
            ybar,
            bar width=6pt,
            xlabel={Number of Nodes},
            ylabel={Workflow Count},
            xtick=data,
            xticklabel style={rotate=45},
            symbolic x coords={0,1,2,3,4,5,6,7,8,9,10,11,12,13,14,16},
            ymin=0,
            ymax=16500,
            nodes near coords,
            nodes near coords align={vertical},
            enlarge x limits=0.05,
        ]
        \addplot+[
            fill=blue
        ] coordinates {
            (0,20) (1,14508) (2,2252) (3,4496) (4,1166) (5,476)
            (6,226) (7,103) (8,143) (9,51) (10,5) (11,28)
            (12,2) (13,33) (14,1) (16,11)
        };
        \end{axis}
    \end{tikzpicture}
    \caption{Distribution of workflows by number of atomic operations.}
    \label{fig:multi_node}
\end{figure}
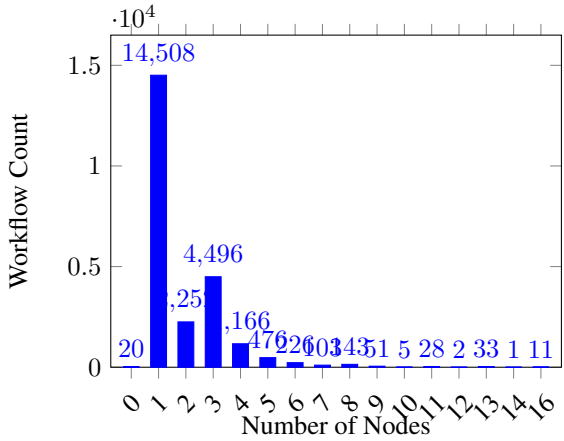

Figure~\ref{fig:multi_node} presents the distribution of workflows by the number of atomic operation nodes. The majority of workflows (14,508) contain only a single atomic step, suggesting that many workflows are used for simple, isolated tasks. However, a significant number of workflows exhibit more complex structures: 2,252 workflows contain two atomic operations, 4,496 contain three, and 1,166 contain four. Additionally, while rarer, workflows with deeper composition (up to 16 atomic nodes) are also observed. These long-tail distributions reflect a wide range of compositional complexity, which imposes challenges on static workflow orchestration and demands adaptive planning capabilities from automated agents.

Such distributional characteristics indicate that real-world workflow automation systems must handle a heterogeneous mixture of simple and highly composed task structures. Static or rule-based agents often fail to generalize across this spectrum, highlighting the need for structure-aware and scalable orchestration methods.

\subsection{Nested Workflow Distribution}

In addition to analyzing the number of atomic operation nodes, we investigate the depth of structural nesting within enterprise workflows, which captures hierarchical dependencies and control flow complexity. From a total of 23,521 workflows, we observe that 7,087 workflows contain nested structures, while the remaining 16,434 are flat.

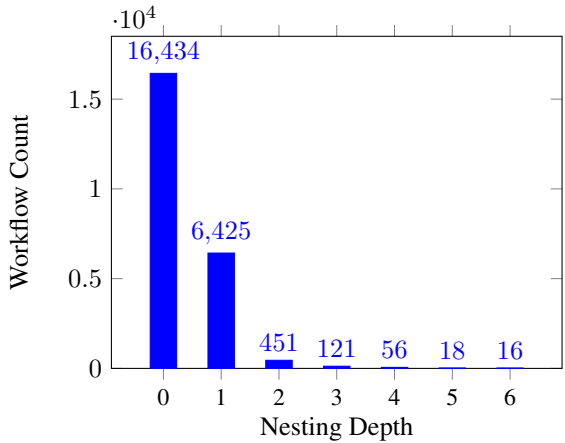
\begin{figure}[ht]
    \centering
    \begin{tikzpicture}
        \begin{axis}[
            width=0.9\linewidth,
            height=6cm,
            ybar,
            bar width=10pt,
            xlabel={Nesting Depth},
            ylabel={Workflow Count},
            xtick=data,
            xticklabel style={font=\small},
            symbolic x coords={0,1,2,3,4,5,6},
            ymin=0,
            ymax=18500,
            nodes near coords,
            nodes near coords align={vertical},
            enlarge x limits=0.15,
        ]
        \addplot+[
            fill=blue
        ] coordinates {
            (0,16434) (1,6425) (2,451) (3,121) (4,56)
            (5,18) (6,16)
        };
        \end{axis}
    \end{tikzpicture}
    \caption{Distribution of workflows by nesting depth.}
    \label{fig:nesting_depth}
\end{figure}

As illustrated in Figure~\ref{fig:nesting_depth}, most nested workflows exhibit shallow nesting: 6,425 workflows have a nesting depth of 1, 451 workflows reach a depth of 2, and only 121 workflows exceed a depth of 2. Deep nesting (depth $\geq$ 3) is rare but not negligible: we identify 56 workflows with a depth of 4, 18 with depth 5, and 16 workflows reaching depth 6.

This skewed distribution highlights that while most workflows follow a relatively flat structure, a considerable number exhibit hierarchical organization, such as conditional branches, loops, or composite sub-tasks. These nested constructs pose challenges for static workflow models and require compositional reasoning capabilities to accurately interpret and execute them.

Such structural complexity necessitates workflow agents that not only identify the atomic operations but also model the recursive and conditional dependencies between them. Our proposed system addresses this by enabling symmetry-guided hierarchical representation, which collectively enhance the agent’s ability to generalize across workflows with varying nesting depths.

%% file: section-appendix-supplementary-experiments.tex
\section{Supplementary Experiments: Atomic Workflow Reuse Efficiency}
We further investigate how different paradigms reuse atomic-level process patterns—defined as reusable subflows composed of 2–5 tool invocations with known semantics (e.g., parse → validate → transform). We measure reuse efficiency as the proportion of workflows that correctly identify and leverage such known subflows during planning.

As shown in Table~\ref{tab:reuse-efficiency}, Structured models (especially SOAN) demonstrate significantly higher reuse efficiency across all complexity levels. SOAN achieves over 72\% reuse in workflows with more than 30 tools, effectively compressing long processes by detecting symmetric or equivalent atomic structures. In contrast, CoT methods lack structural abstraction capabilities and reuse less than 20\% of known subflows. Multi-Agent approaches reuse slightly more (around 25–35\%) by leveraging hard-coded agent templates, but still fall short due to their inability to generalize beyond predefined strategies.

\begin{table}[h]
    \centering
    \begin{tabular}{l|ccc|c}
        \toprule
        \textbf{Method} & \textbf{10--20} & \textbf{20--30} & \textbf{30--40} & \textbf{Avg.} \\
        \midrule
        CoT       & 21.3 & 15.7 & 12.2 & 16.4 \\
        Cot-SC           & 24.5 & 18.9 & 14.3 & 19.2 \\
        AutoGen          & 28.1 & 25.2 & 22.0 & 25.1 \\
        Camel            & 32.0 & 29.6 & 24.4 & 28.7 \\
        MetaGPT          & 29.3 & 26.8 & 23.1 & 26.4 \\
        AFlow            & 58.4 & 49.7 & 42.9 & 50.3 \\
        \textbf{SOAN}    & \textbf{73.1} & \textbf{69.5} & \textbf{65.2} & \textbf{69.3} \\
        \bottomrule
    \end{tabular}
    \caption{Atomic Subflow Reuse Efficiency (\%) across Tool Counts}
    \label{tab:reuse-efficiency}
\end{table}
These results confirm that structural workflow representations, particularly those with symmetry-guided optimization as in SOAN, not only improve planning accuracy but also enhance subflow compression and reuse—key to scaling in large-tool environments.

\section{Supplementary Experiments: Ablation Study on Goal Model Components}
To evaluate the contribution of goal modeling in structured workflow planning, we conduct an ablation study by selectively removing key components of the goal model:

\begin{itemize}
    \item \textbf{w/o Input Goal}: The system receives no structured input specification (i.e., missing field constraints and input semantics), and only uses a generic task description.
    \item \textbf{w/o Output Goal}: The system plans the workflow without an explicit output schema or verification signal, relying only on intermediate tool descriptions.
\end{itemize}

\begin{table}[h]
    \centering
    \begin{tabular}{l|cc}
        \toprule
        \textbf{Variant} & \textbf{Accuracy (\%)} & \textbf{Reuse Rate (\%)} \\
        \midrule
        w/o Input Goal   & 68.4 & 42.3 \\
        w/o Output Goal  & 65.9 & 38.1 \\
        \textbf{Full Model (SOAN)} & \textbf{84.6} & \textbf{69.3} \\
        \bottomrule
    \end{tabular}
    \caption{Ablation Study on Input and Output Goal Modeling}
    \label{tab:goal-ablation}
\end{table}

We evaluate both variants on a subset of high-complexity workflows (tools $\geq $ 30) and report accuracy, plan reuse, and success rate in Table~\ref{tab:goal-ablation}. Results show that both input and output goals are critical: removing the input goal decreases task accuracy due to misaligned tool usage, while removing the output goal harms the system's ability to prune redundant steps and converge to valid outputs. The full model with dual goal anchoring performs best across all metrics.

These findings confirm that both input and output goal anchoring are essential for achieving high-fidelity workflow synthesis in structurally guided frameworks like SOAN.